\documentclass[aps,pre,reprint,superscriptaddress,showpacs]{revtex4-1}
\usepackage{graphicx}
\usepackage{color}
\usepackage{longtable}

\begin{document}

\title{Scaling of cluster heterogeneity in percolation transitions}
\author{Jae Dong Noh}
\affiliation{Department of Physics, University of
Seoul, Seoul 130-743, Republic of Korea}
\affiliation{School of Physics, Korea Institute for Advanced Study,
Seoul 130-722, Republic of Korea}
\author{Hyun Keun Lee}
\affiliation{Department of Physics, University of
Seoul, Seoul 130-743, Republic of Korea}
\author{Hyunggyu Park}
\affiliation{School of Physics, Korea Institute for Advanced Study,
Seoul 130-722, Republic of Korea}

\date{\today}

\begin{abstract}
We investigate a critical scaling law for the cluster
heterogeneity $H$ in site and bond percolations in $d$-dimensional lattices
with $d=2,\cdots,6$. The cluster heterogeneity is defined as the number of
distinct cluster sizes. As an occupation probability $p$ increases, the cluster
size distribution evolves from a monodisperse distribution to a polydisperse
one in the subcritical phase, and back to a monodisperse one
in the supercritical phase. We show analytically that $H$ diverges
algebraically approaching the percolation critical point $p_c$ as $H\sim
|p-p_c|^{-1/\sigma}$ with the critical exponent $\sigma$ associated with the
characteristic cluster size. Interestingly, its finite-size-scaling
behavior is governed by a new exponent $\nu_H = (1+d_f/d)\nu$
where $d_f$ is the fractal dimension of the critical percolating cluster
and $\nu$ is the correlation length exponent. The corresponding scaling variable
defines a singular path to the critical point. All results are confirmed by numerical simulations.
\end{abstract}
\maketitle

\section{Introduction}\label{introduction}
Percolation is a geometric
phase transition for connectivity~\cite{Stauffer,Christensen}.
Suppose that a
fraction $p$ of sites or bonds are occupied in an infinite lattice.
When $p$ is less than a
threshold $p_c$, any site is connected to others up to a finite distance
via occupied sites or bonds.
As $p$ increases, clusters (sets of connected sites) grow
until there emerges a spanning, giant, or an infinite cluster
to which a finite fraction of sites belong.

The percolation transition is manifested in several quantities. The order
parameter $m$, given by the fraction of sites belonging to the giant
cluster, becomes nonzero beyond the transition.
The mean cluster size $S$, defined as the average size of finite clusters
which includes a randomly-selected site, diverges at the transition.
Their scaling properties, such as the exact critical
exponents in two dimension~\cite{exact_result}
and the upper critical dimensionality $d_u=6$~\cite{Stauffer},
are well understood.

Recently, the cluster heterogeneity $H$, defined as the number of distinct
cluster sizes, is suggested as a useful indicator
of a percolation transition~\cite{Lee11}. Consider a cluster size
distribution function $n_s$ which is defined as the number of clusters of
size $s$ per site.
When $p=0$, the distribution is monodisperse with $H=1$.
As $p$ increases, clusters nucleate and aggregate
into larger ones. Consequently, in the subcritical phase,
the cluster size distribution becomes broader.
In the supercritical phase, finite clusters are absorbed
into the giant cluster to decrease $H$.
Hence one expects that the cluster heterogeneity may be maximal
at the transition.

Lee {\it et al.}~\cite{Lee11} found that the maximum
heterogeneity points indeed converge to a percolation threshold in
the thermodynamic limit in the study
of the so-called explosive percolation~\cite{Achlioptas09}.
Exploiting the finite-size-scaling (FSS) property at the maximum heterogeneity points, they
could clarify the nature of the explosive percolation transition.
However, the scaling property of $H$ by itself has not
been understood fully even in the ordinary random percolations.
In this paper, we establish the critical scaling law for
$H$ with emphasis on its FSS theory.

We begin with a brief review for the scaling theory.
For detailed review, we refer readers to Ref.~\cite{Stauffer,Christensen}.
The ordinary percolation exhibits a continuous transition~\cite{comment}.
The correlation length diverges algebraically
as $\xi \sim |\epsilon|^{-\nu}$ where $\epsilon\equiv p-p_c$ and $\nu$ is
the correlation length exponent. At $p=p_c$, the giant cluster is a fractal
characterized with a fractal dimension $d_f$. Physical quantities have a
singular dependence on $\epsilon$. For example, the order parameter
scales as $m \sim \epsilon^\beta$ for
$p\ge p_c$ with the order parameter exponent $\beta$.
The mean cluster size $S$ diverges as $S \sim |\epsilon|^{-\gamma}$
with the susceptibility exponent $\gamma$.
The cluster size distribution function $n_s(p)$ for finite $s$ scales as
\begin{equation}\label{n_s_offcritical}
n_s(p) \sim s^{-\tau} e^{-s/s_c}
\end{equation}
where $\tau$ is called the Fisher exponent and
$s_c$ is the characteristic cluster
size. It diverges as $s_c \sim |\epsilon|^{-1/\sigma}$ with a critical
exponent $\sigma$. There exist scaling relations among those critical
exponents. So, any exponent can be written in terms of two
independent ones, say $\nu$ and $d_f$. The scaling relations read
$\beta = \nu (d-d_f)$, $\gamma = \nu (2 d_f-d)$,
$\sigma = {1}/(\nu d_f)$, and $\tau = 1+{d}/{d_f}$.

In finite systems of linear size $L$, the scaling laws are
modified because the correlation length $\xi$ is limited by $L$.
The FSS hypothesis assumes that a finite-size effect comes into play
through the ratio between $L$ and $\xi$~\cite{Privman}.
It leads to the FSS ansatz for an observable $Q(p,L)$ as
\begin{equation}\label{fss_Q}
Q(p,L) = L^{X_Q} \mathcal{F}_Q(\epsilon L^{1/\nu}) \ ,
\end{equation}
where $X_Q$ is a scaling exponent and $\mathcal{F}_Q(x)$ is a scaling
function for $Q$~\cite{Privman}.
The FSS ansatz claims that a finite system suffers from a finite-size effect
in the region $|\epsilon|<L^{-1/\nu}$ whereas it behaves as an infinite one
elsewhere.
The aim of this paper is to find a scaling law for $H$.

This paper is organized as follows:
In Sec.~\ref{2D_simulation}, we present numerical data for $H$ in site and
bond percolations in two-dimensional (2D) square and triangular lattices, which shows that the
standard FSS form~(\ref{fss_Q}) is not valid for $H$.
In Sec.~\ref{scaling_theory}, we derive the central result of
Eq.~(\ref{H_scaling}) and present numerical data
in $d=2,\cdots,6$ dimensional hypercubic lattices to confirm it.
We summarize and conclude this paper in Sec.~\ref{sec:summary}.

\section{Percolations in 2D} \label{2D_simulation}
\begin{figure}[t]
\includegraphics*[width=\columnwidth]{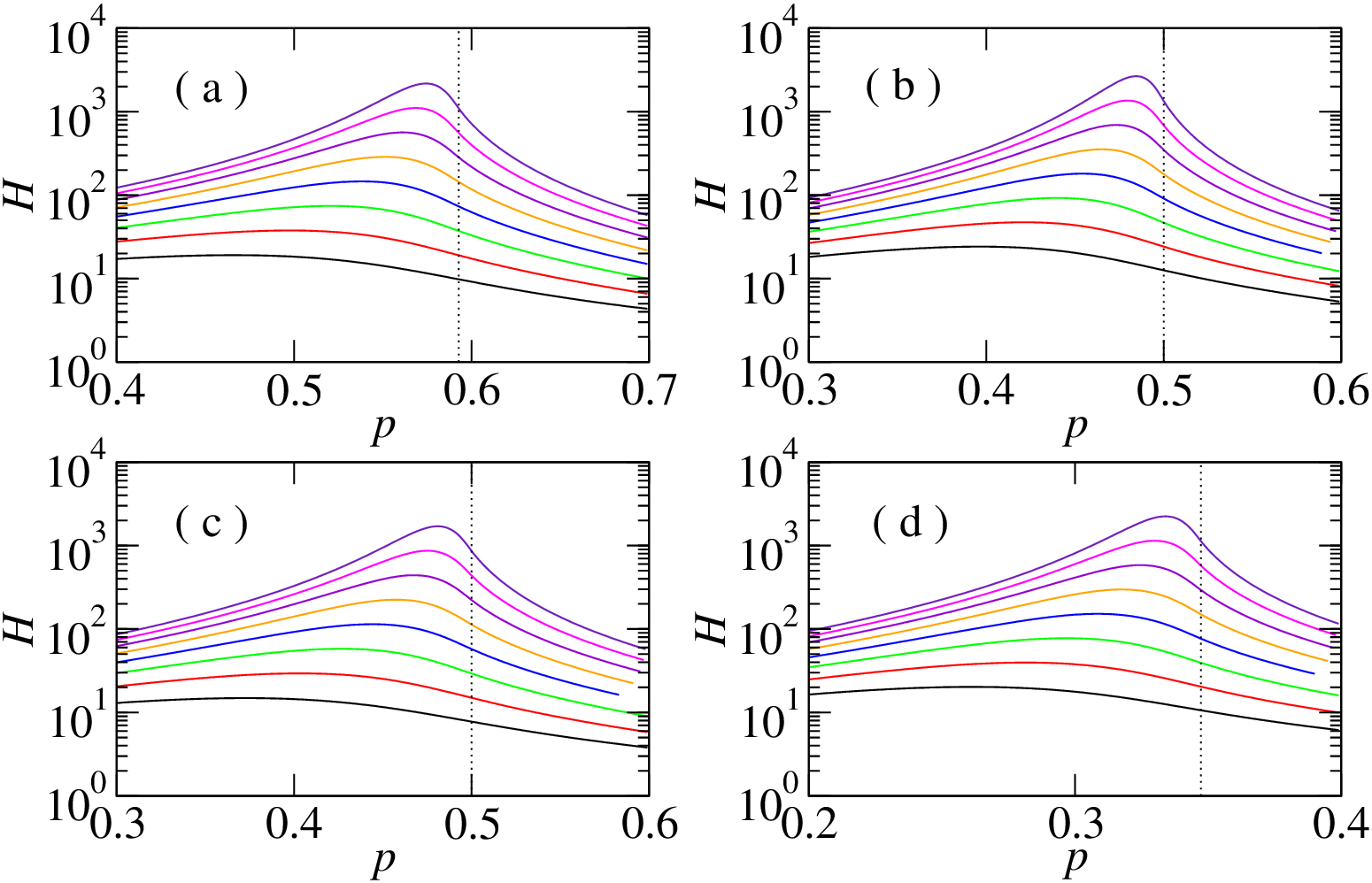}
\caption{(Color online) Cluster heterogeneity in site percolations in
(a) and (c), and bond percolations in (b) and (d) on 2D square lattices in
(a) and (b) and triangular lattices in (c) and (d).
The dotted lines represent the critical percolation threshold.
Lattice sizes are $L=2^5,\cdots,2^{12}$ for the
square lattices and $L=27 \times 2^0, \cdots, 27\times 2^7$ for the
triangular lattices. The larger $L$ is, the higher $H$ is.}
\label{fig1}
\end{figure}
We have performed Monte Carlo simulations of site and bond percolations
in square and triangular lattices of $L\times L$ sites in 2D
using the Newman and Ziff algorithm~\cite{Newman00}. This algorithm allows
an efficient and fast measurement of a quantity $Q$ as a function
of the number $n$ of occupied sites or bonds. A corresponding quantity as a
function of the occupation probability $p$ is then obtained by the
convolution
$Q(p) = \sum_n B(N,n,p) Q(n)$ where $N$ is the total number of sites or
bonds and $B(N,n,p) \equiv \frac{N!}{n!(N-n)!}p^n(1-p)^{N-n}$ is
the binomial distribution function~\cite{Newman00,Hu92}.

\begin{figure}[t]
\includegraphics*[width=\columnwidth]{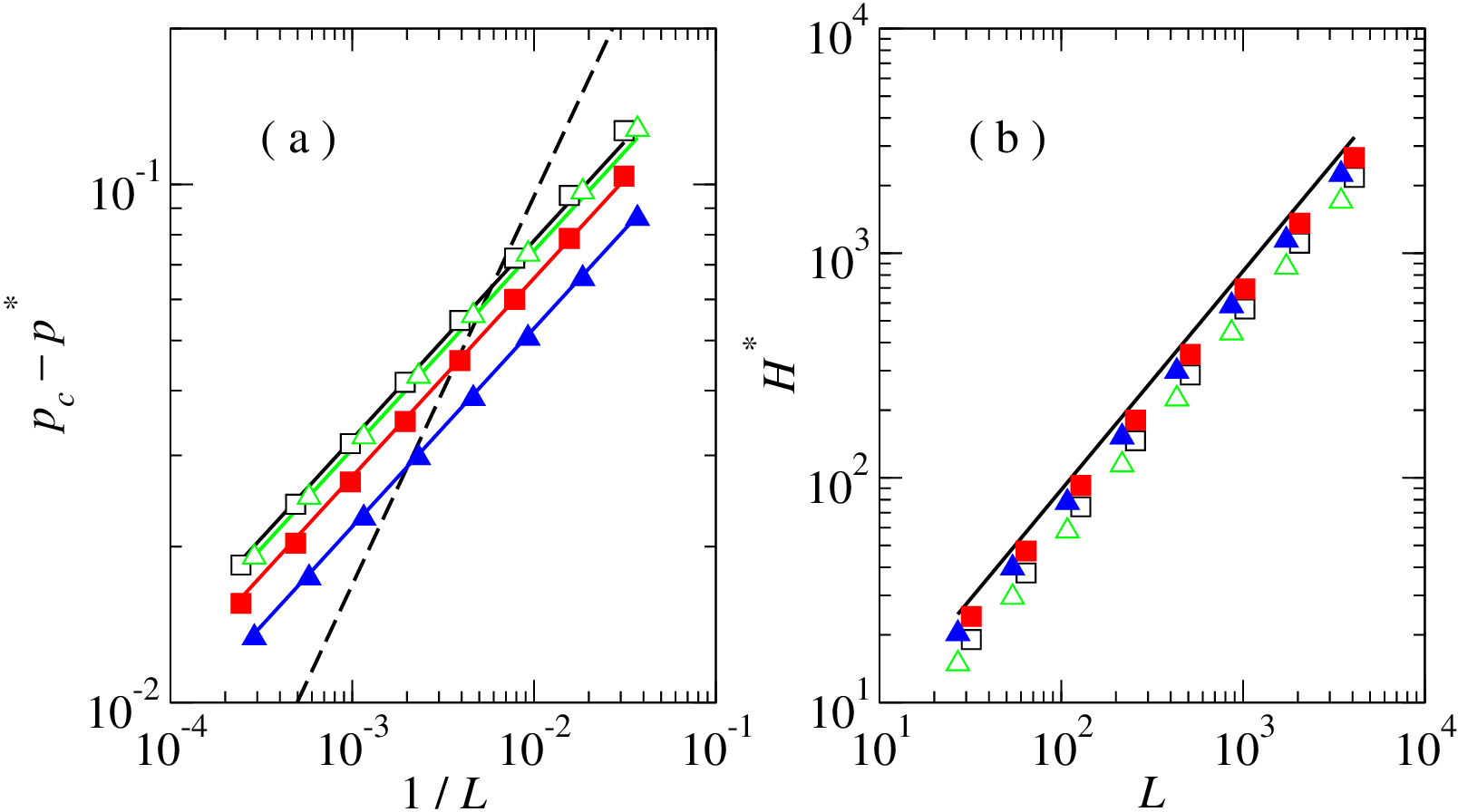}
\caption{(Color online) FSS of $(p_c -p^*)$ in (a) and $H^*$ in (b)
for the site (open symbols) and the bond~(filled symbols) percolations on
2D square~(square symbols) and
triangular~(triangle symbols) lattices.
All straight lines are guides to eyes.
In (a), all the solid lines have the same slope while the dashed line
has a different slope of $1/\nu=3/4$.}
\label{fig2}
\end{figure}

The cluster heterogeneity, averaged over $N_S = 10^5$ samples, is presented
in Fig.~\ref{fig1}.
In all cases, the curves have a diverging
peak (maximum cluster heterogeneity) at a position
denoted by $(p^*, H^*)$. The peak position seems to approach from below
the critical point, which is $p_c=0.592746$ for the square lattice site
percolation~\cite{Newman00}, $1/2$ for the square lattice bond percolation
and the triangular lattice site percolation, and $2\sin(\pi/18)$ for the
triangular lattice bond percolation~\cite{Sykes64}.
FSS properties are analyzed in Fig.~\ref{fig2}.
We find that $(p_c-p^*)$ and $H^*$ scale algebraically with $L$
with apparently universal exponents.

It is noteworthy that $(p_c-p^*)$ does not scale as $L^{-1/\nu}$ with the
correlation length exponent $\nu=4/3$ in $2$D. Such a scaling
would be natural from the FSS hypothesis of Eq.~(\ref{fss_Q}). However, Fig.~\ref{fig2}(a)
excludes such a possibility definitely. It calls for an appropriate FSS theory
for $H$.

\section{FSS theory for $H$}\label{scaling_theory}
In order to characterize the cluster heterogeneity, one needs to specify the
size of each cluster.
Jan {\it et al.}~\cite{Jan98} considered a FSS behavior of the $r$th largest
cluster size at the critical point where the
cluster size distribution follows a power law $n_s \sim s^{-\tau}$.
The $r$th cluster size $s_r$ is estimated from the relation
\begin{equation}
r \sim L^{d} \int_{s_r} ds ~ s^{-\tau} \ ,
\end{equation}
which yields that
\begin{equation}
s_r \sim r^{-\frac{1}{\tau-1}} L^{\frac{d}{\tau-1}} =
r^{-\frac{1}{\tau-1}} L^{d_f} \ .
\end{equation}
We extend their idea to the off-critical region to derive the FSS theory
for $H$.

For $p\lesssim p_c$, the cluster size distribution function is given by
Eq.~(\ref{n_s_offcritical})~\cite{comment1}. Then, the size of a $r$th largest
cluster is obtained from
\begin{equation}\label{r_sr}
r = L^d \int_{s_r}ds~ n_s \sim L^d s_c^{1-\tau}~
\Gamma(1-\tau,s_r/s_c) \ ,
\end{equation}
where function $\Gamma(u,x) \equiv \int_{x}^\infty dt~ t^{u-1} e^{-t}$ is
the incomplete gamma function.
The characteristic size itself displays a FSS behavior~\cite{Christensen}
\begin{equation}\label{s_cutoff}
s_c \sim \left\{ \begin{array}{ccc}
   |\epsilon|^{-1/\sigma} &\mbox{for}& |\epsilon| \gg L^{-1/\nu}\ , \\[2mm]
   L^{d_f}                  &\mbox{for}& |\epsilon| \ll L^{-1/\nu} \ .
\end{array}\right.
\end{equation}

We now compare the average size of $r$th and $(r+1)$th clusters.
Using Eq.~(\ref{r_sr}), one finds that
\begin{equation}\label{Delta_sr}
\Delta_r s = s_{r}-s_{r+1} \sim L^{-d} s^{\tau} e^{s/s_c} \ .
\end{equation}
The distribution is {\em dense} when $\Delta_r s<1$ and {\em
sparse} when $\Delta_r s>1$.
The two regions are separated at $s=s_0$ satisfying
\begin{equation}\label{crossover}
L^{-d} s_0^{\tau} e^{s_0/s_c} = \mathcal{O}(1)  \ .
\end{equation}
The rank of a cluster of size $s_0$ is given by
\begin{equation}\label{r0}
r_0 \sim L^d s_c^{1-\tau} ~\Gamma(1-\tau,s_0/s_c) \ .
\end{equation}
Note that $L^d n_{s_0} = \mathcal{O}(1)$.
This implies that there are at least $\mathcal{O}(1)$ clusters of
all sizes $s < s_0$. On the other hand, there are $r_0$ clusters in the
region $s>s_0$, whose sizes are distinct because $\Delta_r s>1$.
Therefore we find that
\begin{equation}
H \simeq s_0 + r_0  \ .
\end{equation}

To obtain the solution of Eq.~(\ref{crossover}) for $s_0$, we first assume that
$s_0 \ll s_c$. Then, the exponential term is negligible and the solution is
given by
\begin{equation}\label{s_0_critical}
s_0 \sim L^{d/\tau} \ .
\end{equation}
Using the asymptotic behavior $\Gamma(u,x \to 0) \sim -\frac{1}{u} x^{u}$,
one also finds that $r_0 \sim s_0$.
This solution is self-consistent when
$L^{d/\tau} \ll |\epsilon|^{-1/\sigma}$ or
\begin{equation}
|\epsilon| \ll L^{-1/\nu_H}
\end{equation}
with the exponent $\nu_H$
\begin{equation}\label{nu_H}
\nu_H = \frac{\tau}{d\sigma} =
\frac{\tau}{\tau-1}\nu = \left(1+\frac{d_f}{d}\right) \nu \ .
\end{equation}
In the opposite case $s_0 \gg s_c$, the leading order
solution of Eq.~(\ref{crossover}) is given by
\begin{equation}\label{s_0_off}
s_0 \sim s_c \ln L \ .
\end{equation}
From the asymptotic behavior $\Gamma(u,x\to\infty) \sim x^{u-1}e^{-x}$, one
also finds that $r_0 \sim s_c$. This solution is self-consistent
when $|\epsilon| \gg L^{-1/\nu_H}$.

Using the solution for $s_0$ and $r_0$, we can summarize the
scaling property of $H$
with the FSS form
\begin{equation} \label{H_scaling}
H(p,L) = L^{{d}/\tau }
\mathcal{F}_H(\epsilon L^{1/\nu_H}) \ .
\end{equation}
The scaling function has a limiting behavior
\begin{equation}
\mathcal{F}_H(x) \sim \left\{ \begin{array}{ccc}
|x|^{-1/\sigma} \ln |x| &\mbox{for}& |x| \gg 1 , \\[2mm]
\mbox{constant} &\mbox{for}& |x| \ll 1
\end{array}\right.
\end{equation}
so that $H \sim (\ln L) |\epsilon|^{-1/\sigma}$ for $|\epsilon|\gg
L^{-1/\nu_H}$ and $H \sim L^{d/\tau}$ for $|\epsilon|\ll L^{-1/\nu_H}$.

\begin{figure}[t]
\includegraphics*[width=\columnwidth]{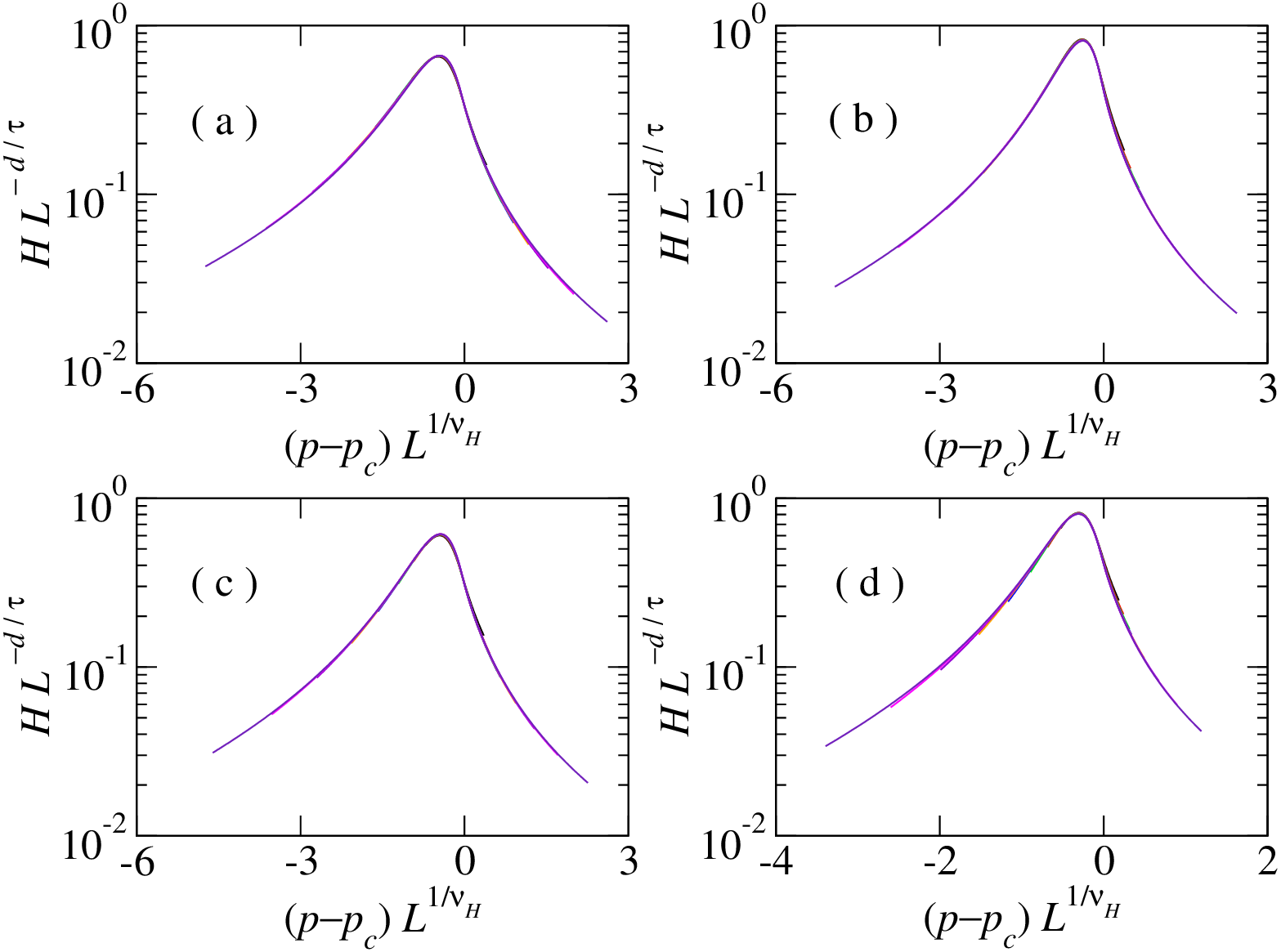}
\caption{(Color online) Scaling analysis of $H$ for
the site ((a) and (c)) and bond ((b) and (d)) percolations on
2D square ((a) and (b)) and triangular ((c) and (d)) lattices.
The same data sets are used as in Fig.~\ref{fig1}.}
\label{fig3}
\end{figure}

Remarkably, the FSS exponent $\nu_H$ for the
cluster heterogeneity is distinct from the correlation length exponent $\nu$.
It explains the reason why $(p_c-p^*)$ does not scales as
$L^{-1/\nu}$ in Fig.~\ref{fig2}(a). Instead, the numerical data are
consistent with the scaling $(p_c-p^*) \sim L^{-1/\nu_H}$ with $\nu_H =
187/72$ in 2D~(see Table~\ref{table1}).
The solid lines in Fig.~\ref{fig2}(a) have the slope of $1/\nu_H$.
The maximum heterogeneity is expected to scale as $H^* \sim L^{d/\tau}$
with $d/\tau= 182/187$ in 2D.
The solid line in Fig.~\ref{fig2}(b) is of this slope and in agreement
with the numerical data.
The FSS scaling form is tested in Fig.~\ref{fig3},
where we replot the data in Fig.~\ref{fig1}
according to Eq.~(\ref{H_scaling}).
All the data from different sizes collapse perfectly onto a single curve.

\begin{table*}
\caption{Critical points and critical exponents in the site
percolations in $d$-dimensional hypercubic lattices.}\label{table1}
\begin{ruledtabular}
\begin{tabular}{cccccc}
  & $d=2$ & $d=3$ & $d=4$ & $d=5$ & $d=6$ \\ \hline
$p_c$ (site) & $0.592746$~\cite{Newman00}  & $0.3116077$~\cite{Deng05} & $0.196889$~\cite{Paul01} &
$0.140765$~\cite{Grassberger03} & $0.109017$~\cite{Grassberger03} \\
$\nu$ &  $4/3$   & $0.875$~\cite{Lorenz98} & $0.689$~\cite{Ballesteros97} &
$0.51$~\cite{Jan85} & $1/2$~(MF) \\
$d_f$ &  $91/48$ & $2.523$~\cite{Lorenz98} & $3.05$~\cite{Paul01} &
$3.54$~\cite{Paul01} & $4$~(MF) \\
$\nu_H = (1+d_f/d)\nu$ & $187/72$ & $1.611$ & $1.21$ & $0.87$ & $5/6$~(MF) \\
$d/\tau = d/(1+d/d_f)$ & $182/187$ & $1.370$ & $1.73$ & $2.07$ & $12/5$~(MF) \\
\end{tabular}
\end{ruledtabular}
\end{table*}

We examine whether the FSS form in Eq.~(\ref{H_scaling}) for $H$
is valid universally in
higher dimensions up to the upper critical dimension $d_u=6$.
We have performed extensive numerical simulations of site
and bond percolations to measure $H$
in hypercubic lattices in $d=2,\cdots,6$ dimensions.
The system sizes are $L=2^3,\cdots, 2^8$ in 3D, $L=2^3,\cdots, L^6$ in
4D, $L=8,\cdots,28$ with $\Delta L = 4$ for 5D,
and $L=6,\cdots,16$ with $\Delta L = 2$ for 6D.

\begin{figure}[t]
\includegraphics*[width=\columnwidth]{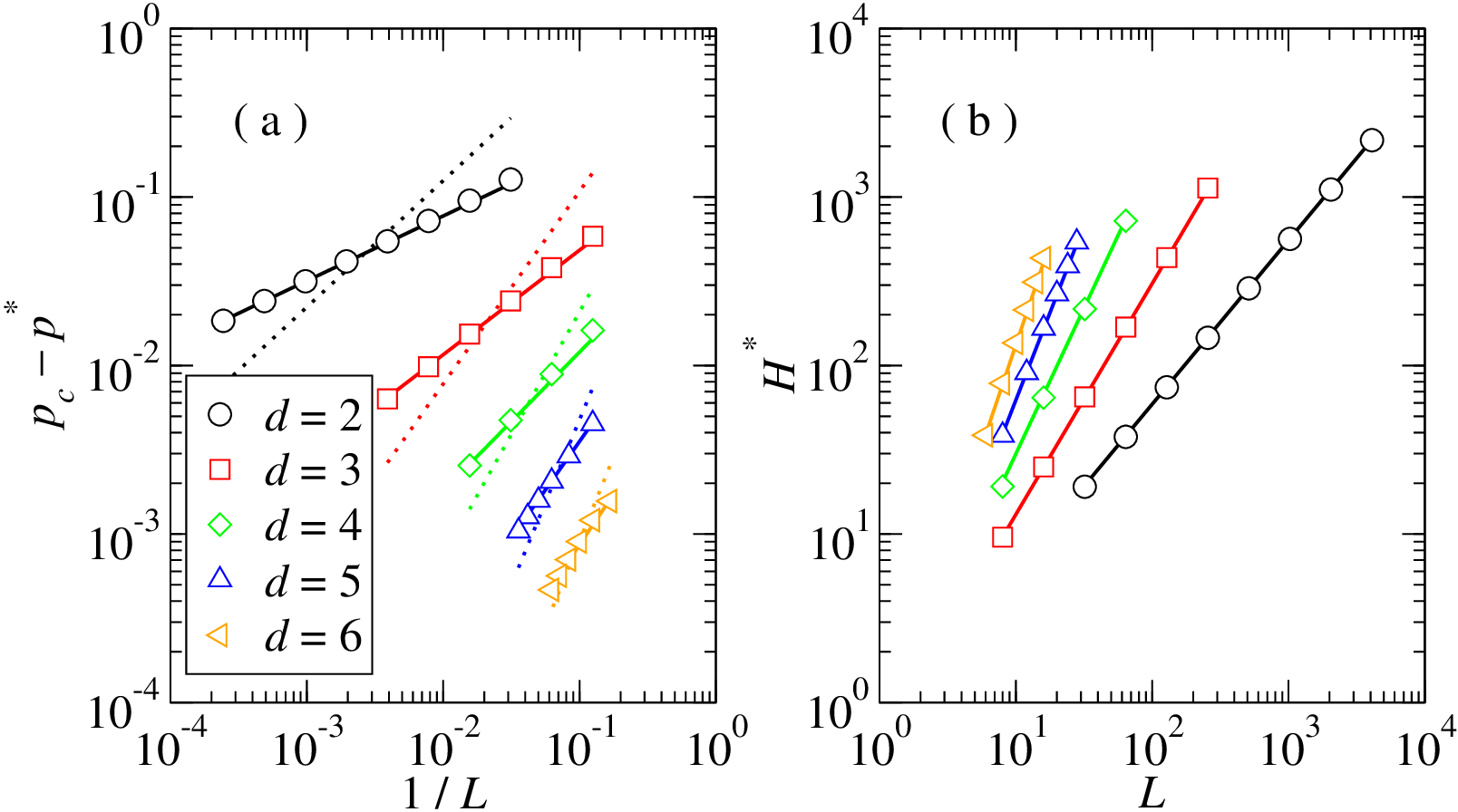}
\caption{(Color online) (a) $(p_c-p^*)$ vs. $1/L$. The solid lines have a slope of
$1/\nu_H$ while the dotted lines have a slope of $1/\nu$.
(b) $H^*$ vs. $L$ with the solid lines having a slope of $d/\tau$.}
\label{fig4}
\end{figure}

In Fig.~\ref{fig4}(a), $(p_c-p^*)$ is plotted against $1/L$ for the site
percolations. Also drawn are the straight lines corresponding to the
scalings $L^{1/\nu}$ and $L^{1/\nu_H}$.
The parameter values used in the plot are collected from literatures and
listed in Table~\ref{table1}. In all dimensions, $(p_c-p^*)$ scales
with the exponent $1/\nu_H$ instead of $1/\nu$. Figure~\ref{fig4}(b) shows
the plot of $H^*$ against $L$, which also confirms the
scaling $H^*\sim L^{d/\tau}$. We have obtained the same result
for the bond percolations, whose data are not shown here.

\section{Summary and Discussion}\label{sec:summary}
We have shown that the cluster heterogeneity diverges at the percolation
critical point and that it satisfies the FSS form in Eq.~(\ref{H_scaling}).
The FSS is governed not by the correlation length exponent
$\nu$ but by the new exponent $\nu_H = (1+d_f/d)\nu$.
Consequently, the maximum cluster heterogeneity points
follow the scaling $(p_c-p^*(L)) \sim L^{-1/\nu_H}$.  This is
contrasted with the standard $L^{-1/\nu}$ scaling of the effective critical
points obtained from the order parameter data and the average cluster size data.

Since $\nu_H > \nu$, the maximum heterogeneity points $p^*(L)$ constitute a
singular path of $\epsilon\sim -L^{-1/\nu_H}$ in the subcritical phase to the critical point. Although they converge to $p_c$ in the
$L\to\infty$ limit, the system along the path remains outside the critical
region defined by the condition $|\epsilon| \ll L^{-1/\nu}$.
Physical quantities along a singular path may exhibit a peculiar FSS
behavior.

Consider, for example, the order parameter $m$
which follows the FSS form in Eq.~(\ref{fss_Q}) with $X_m = -\beta/\nu$.
In the subcritical phase, it scales with $L$ as $m\sim L^{-d}$
in the leading order~\cite{Margolina}, which requires that
the scaling function should have a limiting behavior
$\mathcal{F}_m( x\ll -1) \sim |x|^{-(d\nu - \beta)}$.
So the order parameter $m^*$, evaluated
at the maximum heterogeneity points,
follows the scaling $m^* \sim L^{-\beta/\nu}
\mathcal{F}_m(-L^{-1/\nu_H+1/\nu})  \sim L^{-\beta/\nu-d_f (1-\nu/\nu_H)}$,
instead of the standard critical scaling
$m^*\sim L^{-\beta/\nu}$.
In fact, this is a manifestation of the crossover phenomena near a
multicritical point in the context of the FSS~\cite{Fisher}. When there
exist multiple relevant scaling fields for a critical point,
scaling behaviors become dependent on a path approaching it, in general.
Since $1/L$ is one of such relevant scaling
variables~\cite{Nightingale,Privman}, one can naturally
expect a nontrivial FSS along a singular path.
The maximum cluster heterogeneity condition indeed allows us to access a singular path to the
percolation critical point.

This work was supported by Mid-career Researcher Program through NRF grant
(No.~2011-0017982) funded by the MEST.


\begin{thebibliography}{99}
\bibitem{comment} There has been a controversy whether there exists a
        discontinuous percolation transition since the paper in
        Ref.~\cite{Achlioptas09} was published. For example, see Ref.~\cite{Lee11}
        and references therein for ongoing debates on the explosive percolation transitions.
\bibitem{comment1} In the supercritical phase, $n_s$ has an additional
isolated peak representing the giant cluster. It does not modify the scaling
behavior of $H$.
\end{thebibliography}
\end{document}